\documentclass[amsmath,amssymb,aps,prb,superscriptaddress,twocolumn, fleqn]{revtex4}
\usepackage{graphicx}
\usepackage{dcolumn}
\usepackage{bm}
\usepackage{xcolor}
\usepackage{afterpage}

\preprint{APS/123-QED}

\begin{document}
\title{Magnetodielectric Properties in Two Dimensional Magnetic Insulators}
\author{Koushik Dey}
\thanks{koushikdey664@gmail.com}
 \affiliation{Technical Research Center (TRC), Indian Association for the Cultivation of Science (IACS), Jadavpur, Kolkata 700032, India}
 \affiliation{School of Physical Sciences, Indian Association for the Cultivation of Science (IACS), Jadavpur, Kolkata 700032, India}
 \author{Hasina Khatun}
  \affiliation{School of Physical Sciences, Indian Association for the Cultivation of Science (IACS), Jadavpur, Kolkata 700032, India}
 \author{Anudeepa Ghosh}
  \affiliation{School of Physical Sciences, Indian Association for the Cultivation of Science (IACS), Jadavpur, Kolkata 700032, India}
 \affiliation{Indian Institute of Science Education and Research Kolkata, Mohanpur, Nadia 741246, West Bengal, India}
 \author{Soumik Das}
  \affiliation{School of Physical Sciences, Indian Association for the Cultivation of Science (IACS), Jadavpur, Kolkata 700032, India}
 \author{Bikash Das}
  \affiliation{School of Physical Sciences, Indian Association for the Cultivation of Science (IACS), Jadavpur, Kolkata 700032, India}
 \author{Subhadeep Datta}%
 \thanks{sspsdd@iacs.res.in}
  \affiliation{School of Physical Sciences, Indian Association for the Cultivation of Science (IACS), Jadavpur, Kolkata 700032, India}
 
\begin{abstract}
Magnetodielectric (MD) materials are important for their ability to spin-charge conversion, magnetic field control of electric polarization and vice versa. Among these, two-dimensional (2D) van der Waals (vdW) magnetic materials are of particular interest due to the presence of magnetic anisotropy (MA) originating from the interaction between the magnetic moments and the crystal field. Also, these materials indicate a high degree of stability in the long-range spin order and may be described using suitable spin Hamiltonians of the Heisenberg, XY, or Ising type. Recent reports have suggested effective interactions between magnetization and electric polarization in 2D magnets. However, MD coupling studies on layered magnetic materials are still few. This review covers the fundamentals of magnetodielectric coupling by explaining related key terms. It includes the necessary conditions for having this coupling and sheds light on the possible physical mechanisms behind this coupling starting from phenomenological descriptions. Apart from that, this review classifies 2D magnetic materials into several categories for reaching out each and every class of materials. Additionally, this review summarizes recent advancements of some pioneer 2D magnetodielectric materials. Last but not the least, the current review provides possible research directions for enhancing magnetodielectric coupling in those and mentions the possibilities for future developments.

\textbf{Keywords:} dielectric spectroscopy, magnetodielectric, multiferroics, ferroelectric

\end{abstract}    
\maketitle

\section{Introduction}
  Novel materials with coupled charge, spin, orbital, lattice degrees of freedom allow in controlling and tuning of functional properties through external field \cite{husain2024non, vaz2024voltage, huang2024manipulating, sun2012voltage}. Among these, coupling between magnetic (spin) and dielectric (charge) properties is a promising avenue for advancing next generation low-power energy-efficient electronics. The spin-charge coupling is considered as more general coupling which do not require any symmetry restrictions or having spontaneous polarization, named as magnetodielectric (MD) coupling \cite{lawes2009magnetodielectric}. Therefore, MD coupling can, in principle be occurred in any insulating magnet which expands the search for potential MD materials and offers greater flexibility in designing novel devices. The spin-charge coupling in MD materials could be utilized in creating innovative non-volatile devices, such as memories, sensors, and spin-charge transducers \cite{luo2024magnetoelectric, kleemann2017multiferroic, bibes2008towards,chikara2019magnetoelectric, padhan2008magnetodielectric}.

To control dielectric properties with the applied magnetic field, two primary strategies have been utilized either by manipulating the interface physics between ferroelectric (FE) and ferromagnetic (FM) heterostructure systems \cite{arora2022unravelling, narayanan2012hybrid, chatterjee2024interfacial} or by enhancing the MD effect nearby magnetic transitions in magnetic insulator systems \cite{kimura2003magnetic, kimura2003magnetocapacitance, lawes2003magnetodielectric, hur2004colossal, tackett2007magnetodielectric, lorenz2004large}. For both the systems, considerable efforts have been devoted to understanding and tailoring MD coupling. However, most studied materials are based on inorganic non-layered oxides and organic/inorganic heterostructures \cite{huang2018spin,huang2022magnetic}. 

 2D layered materials are characterized by weak van der Waals (vdW) forces along the third direction perpendicular to the layers and strong covalent bonding holding atomic structures within 2D layers. Fundamentally, a system's quantum fluctuations are enhanced when its dimensionality is decreased, opening the door to the detection of novel phases and phenomena. Mermin-Wagner theorem, which was formalized in 1966, states that for an isotropic Heisenberg system at finite temperature, long-range magnetic ordering is prohibited in 2D \cite{jenkins2022breaking}. In fact, gapless spin excitations with a finite density of states are produced by the continuous symmetry. This leads to huge thermal fluctuations that suppress any long-range magnetic order \cite{gibertini2019magnetic}. To counteract these fluctuations and stabilize a magnetic order, anisotropy is required to open up a gap in the magnon spectrum \cite{wang2022magnetic}. Despite this, Many 2D materials have been reported to have stable magnetic order down to single atomic layers and the list of these materials are growing rapidly \cite{tang2019two}. Recently, there has been a surge in research for studying magnetic and electric coupling phenomena on 2D vdW materials for their exceptional application prospects. However, MD studies on 2D magnetic insulators are still at an infant stage.   

Recent advancements in 2D vdW magnetic materials offer potential to revolutionize the use of MD materials in multistate storage, spintronics, and nanoelectronics \cite{yang2019design, gu2023multi}. Their atomic scale thickness enables high-density integration on various substrates including silicon. Moreover, free of dangling bonds and charge traps, clean vdW interfaces of these layered materials support easy integration on various other 2D materials \cite{novoselov20162d}. 2D vdW dielectric materials with a larger dielectric constant ($\epsilon$) are often used as gate materials by integrating them onto 2D semiconductor channels. The gate dielectric acts as a voltage amplifier, with amplification proportional to the dielectric constant, thereby enhancing the performance of transistors. Besides the applications, these materials are also interesting for their exotic physical phenomena.      
 
Generally, MD properties are routinely measured through broadband dielectric spectroscopy measurements with applied magnetic fields. This gives detailed insight into abundant underlying fundamental physics such as structural phase transition, charge order, polar nano regions, defects and various relaxation phenomenon \cite{yang2023dielectric}. In 2D materials the similar phenomenon can be investigated by dielectric measurements with applied magnetic fields. The existing review articles on MD properties highlight results on inorganic oxide magnets \cite{lawes2009magnetodielectric, yang2023dielectric, su2014low, bussmann2017unexpected}. Though, there are several recent insightful studies on MD properties of 2D materials, a comprehensive overview of those results is absent. This review comprehensively focusing on coupled MD phenomenon of several 2D materials, highlight key research topics and trends, and identify the possibilities along with potential future developments.

The review will start with the fundamentals of the MD effect in magnetic insulators along with a brief overview of 2D magnetism, followed by interplay between magnetic and dielectric properties in 2D magnetic materials, in section 2 and section 3. The following section will cover the intrinsic and extrinsic origin of MD coupling. In Section 4, the classification of MD materials into subgroups based on different magnetic phenomena will be discussed. This section will also focus on several interesting MD properties measured so far on 2D materials from each subgroup. This review will cover the research topics that further broaden the field of MDs and will discuss on the possibilities to explore on that direction. It will conclude with a summary and an outlook on future developments in Section 5. 


\section{Magnetodielectric coupling}
\label{sec3} 
In typical dielectric measurements, an external electric field is applied in a parallel-plate capacitor which contains the dielectric material under study. The induced polarization in the studied dielectric material is measured. After the application of external electric field $\mathbf{E}$, the elementary dipoles take some time to align along the applied field direction, which results into a gradual change of induced polarization ($\mathbf{P}$) in that direction. 

Magnetism originates from spin degrees of freedom, while the dielectric response is associated with charge. Since electrons have both charge and spin, various mechanisms lead to spin-charge coupling or MD coupling. To understand the nature of coupling between magnetic and charge degrees of freedom, Landau free energy can be expressed as a function of magnetization ($\mathbf{M}$), polarization ($\mathbf{P}$), and the external electric field ($\mathbf{E}$) \cite{lawes2009magnetodielectric, kimura2003magnetocapacitance}, is given below:

\begin{equation}
F = \frac{1}{2\epsilon_0} \mathbf{P}^2 - \mathbf{P} \cdot \mathbf{E} - \alpha (\mathbf{P} \cdot \mathbf{M}) - \beta (\mathbf{P} \cdot \mathbf{M}^2) - \gamma (\mathbf{P}^2 \cdot \mathbf{M}^2) 
\label{free_energy}
\end{equation}

where $\alpha$, $\beta$, and $\gamma$ are coupling constants and $\epsilon_0$ is the free space dielectric permittivity. In this expression, the term $\gamma (\mathbf{P}^2 \cdot \mathbf{M}^2)$ is solely responsible for MD effects. This term is the simplest one involving both $\mathbf{P}$ and $\mathbf{M}$ that produces a scalar for all systems, whatever may be the symmetry of the lattice or magnetic structure \cite{lawes2009magnetodielectric}. This expression has been applied to interpret MD coupling in various non-ferroelectric MD systems as it can describe well the deviation of susceptibility (or dielectric constant) below magnetic transition temperature \cite{kimura2003magnetocapacitance}. 

 The term $\alpha (\mathbf{P} \cdot \mathbf{M})$ represents linear magnetoelectric (ME) coupling. There are 13 groups out of the 122 Shubnikov magnetic point groups allow for simultaneous non-zero polarization and magnetization \cite{schmid2008some}. The term $\beta (\mathbf{P} \cdot \mathbf{M}^2)$ though forbidden by symmetry, is relevant for understanding some ME multiferroics. ME materials are an important class of materials which is extensively discussed in several reviews \cite{fiebig2005revival, zvezdin2004phase, ryu2002magnetoelectric, eerenstein2006multiferroic, nan2008multiferroic}. Current review focuses primarily on MD effects, which are not typically categorized under MEs and multiferroics. 

 Dielectric constant is determined by the second order derivative of $F$ with respect to $P$. Hence, dielectric constant $\epsilon$ is $\epsilon_0$ with an added correction which is proportional to $\gamma \mathbf{M}^2$, using the equation \ref{free_energy}. Therefore, the mentioned free energy expression can be relevant for studying MD coupling in materials having non zero $M$ (\textit{i.e.} FM materials) and can not be described for AFM materials, where net magnetization is zero. To better understand the MD coupling in AFMs, Lawes \textit{et. al} \cite{lawes2009magnetodielectric} introduced a model that couples the uniform polarization \(P\) to the wave vector \(q\)-dependent magnetic correlation function \(\langle M_q M_{-q} \rangle\)  and the modified MD term in the free energy is denoted as \(F_{MD}\), given by:

\begin{equation}
F_{MD} = \sum_{q} g(q) P^2 \langle M_q M_{-q} \rangle
\label{AFM_correction}
\end{equation}

where, g(q) is the q dependent coupling constant and \(\langle M_q M_{-q} \rangle\) is the thermal average of spin spin correlation. The resulting value of polarization is given as \cite{lawes2003magnetodielectric} 
\begin{equation}
P =  \frac{E}{1/\epsilon_0 + 2\epsilon_0 \sum_{\mathbf{q}} g(q) \langle M_{\mathbf{q}} M_{-\mathbf{q}} \rangle (T)} = \epsilon E
\label{polarization}
\end{equation}

where dielectric constant $\epsilon$ is
\begin{equation}
\epsilon =  \frac{\epsilon_0}{1 + 2\epsilon_0 \sum_{\mathbf{q}} g(q) \langle M_{\mathbf{q}} M_{-\mathbf{q}} \rangle (T)}
\label{final_epsilon}
\end{equation}

The above equations \ref{polarization} and \ref{final_epsilon} suggest that the dynamic behavior of spins, such as their correlations and fluctuations, can directly influence the material's polarization and consequently its dielectric properties.


\section{2D magnetism}
\label{sec2}
2D magnetism explores the arrangements of magnetic moments and spin dynamics, specifically within the context of magnetic interactions and spin fluctuations constraints in the two dimensional limit. In these systems, the spin interactions in vdW magnetic materials are often described by a generalized spin Hamiltonian \cite{ahn2024progress}, which approximates the nearest-neighbor interactions as follows:

\begin{equation}
H = -\sum_{\langle i, j \rangle} J_{ij} (S^x_iS^x_j + S^y_iS^y_j + \alpha S^z_iS^z_j) - g\mu_B \sum_{i} \mathbf{B} \cdot \mathbf{S}_i
\label{nn_hamiltonian}
\end{equation}

Here, $J_{ij}$ represents the magnetic exchange coupling between neighboring spins $S^{x,y,z}$, where positive values indicate FM (FM) interactions, and negative values suggest AFM (AFM) interactions. The dimensionless parameter $\alpha$ quantifies the magnetic exchange anisotropy between the in-plane components ($S^x, S^y$) and the out-of-plane component ($S^z$). The last term represents the Zeeman interaction under the external magnetic field $\mathbf{B}$, where g and $\mu_B$ are Lande g-factor and Bohr magneton.

\begin{figure*}
\includegraphics[scale=0.7]{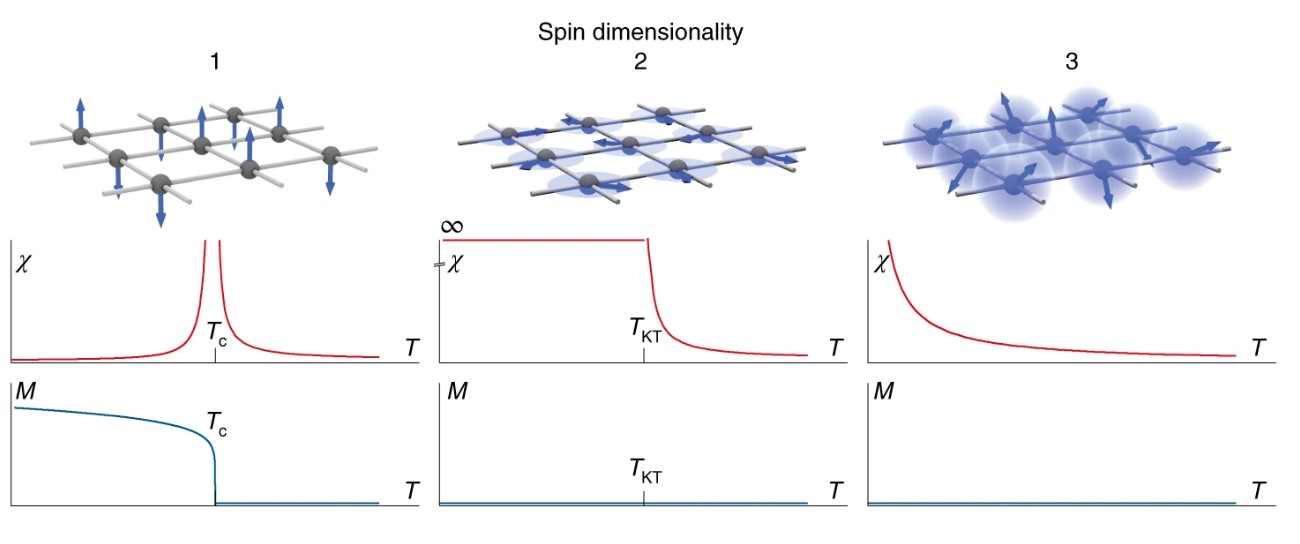}
\caption{(a) In 2D materials, long-range magnetic order develops at a finite temperature depending on the spin dimensionality $n$. For the case $n = 1$, the spins point either up or down along a given direction. The system effectively exhibits a single spin component along the easy axis, and the spin Hamiltonian is described by the Ising model. For $n = 2$, the spins lie in a given plane with effectively two components, i.e., easy-plane anisotropy. The spin Hamiltonian for $n = 2$ is described by the XY model. For this case, $\chi \to \infty$ at $T \to T_{KT}$. For $n = 3$, spins can lie in any direction, and the spin Hamiltonian is described by the isotropic Heisenberg model. Reproduced from \cite{gibertini2019magnetic}. CC BY 4.0.}
\label{dimension}
\end{figure*}

The value of $\alpha$ dictates the type of magnetic anisotropy in the system: when $\alpha > 1$, the system exhibits easy-axis anisotropy, while $\alpha < 1$ corresponds to easy-plane anisotropy. For $\alpha = 1$, the system is isotropic, characterized by Heisenberg model with spin dimentionality $n=3$. In the extreme limit where $\alpha \gg 1$, the system resembles the Ising model, with the spin degree of freedom reduced to one dimension ($n = 1$). For $\alpha = 0$, the system is XY-like, where spins can rotate within the plane, and $n = 2$ (see Fig. \ref{dimension}). 

For 2D vdW materials, due to the strong geometrical anisotropy, significant difference present in the magnitude between the intralayer exchange coupling ($J_{\parallel}$) and interlayer exchange coupling ($J_{\perp}$). This difference leads to a distinction in the properties of these materials when moving from the bulk to the 2D limit \cite{gibertini2019magnetic}. Along with exchange anisotropy, magnetocrystalline anisotropy arising from interactions between magnetic ions and their local crystalline environment, play a crucial role in describing the material's magnetic properties. This can be represented by the the single ion anisotropic \cite{joy1992magnetism} term:

$$- \sum_{i} A_i (S_i^z)^2$$

For chiral spin textures, deviated from collinear FM and AFM alignments, the antisymmetric exchange coupling between neighboring spins, the spin-spin interaction generally presented by the Dzyaloshinskii-Moriya interaction (DMI), expressed as:

$$- \sum_{\langle i, j \rangle} D_{ij} \left( S_i^x S_j^y - S_i^y S_j^x \right),$$

D$_{ij}$ indicates Dzyaloshinskii vector which points toward along a high symmetry direction of the magnetic materials, which is proportional to the spin-orbit coupling constant. 


\subsection{Interplay between magnetism and dielectric properties in 2D materials}
\label{sec2.2}

The interplay between magnetism and dielectric properties in 2D vdW materials is heavily influenced by their intrinsic anisotropies, including both spin dimensionality and lattice. These materials exhibit distinct MD responses due to their low-dimensional nature and the influence of factors such as anisotropic spin-lattice interactions and contrasting intralayer and interlayer exchange couplings. In Ising systems, where the spins are constrained along the out-of-plane direction, a contrast in MD coupling arises along different crystallographic directions \cite{nishiwaki2003dielectric,ghosh2023anisotropic}. On the contrary, XY and Heisenberg-like systems, with their in-plane spin components, give rise to more complex MD coupling \cite{aoyama2017anisotropic,wu2004coupling,sosa2012study}.   

Additionally, the distinction between intralayer and interlayer exchange coupling significantly influences the MD response in 2D vdW materials. Intralayer exchange coupling, which is typically stronger, governs collective spin dynamics within a single layer. The nature and strength of these interactions determine the spin ordering and its coupling to lattice deformations. In such systems, MD coupling often arises from spin-lattice coupling, where electric fields induce lattice distortions that, in turn, modify the spin-charge coupling.

In multilayer vdW materials, interlayer coupling, typically weaker than intralayer coupling, influences the stacking-dependent spin order. In bilayer or multilayer systems, variations in interlayer coupling (whether FM or AFM) can alter the overall magnetic order, providing a pathway to control MD coupling by adjusting the layer stacking or alignment \cite{gibertini2019magnetic, son2021air,huang2017layer}. Additionally, stacking faults between the interlayers of these materials can lead to significant alterations in their overall dielectric properties and magnetodielectric couplings \cite{mi2021stacking,zheng2018dielectric}. Dielectric responses in these multilayer systems can also exhibit anisotropic behavior due to the complex interplay between in-plane and out-of-plane spin-lattice interactions. 

Furthermore, anisotropic spin-lattice interactions can influence the dielectric susceptibility tensor, potentially leading to pronounced magnetocapacitance effects, particularly near or below magnetic phase transitions. \cite{paul2015spin,chu2015real}. The magnitude and tunability of the dielectric response in these materials depend on the direction of external magnetic fields and electric fields \cite{aoyama2017anisotropic}. In the presence of external electric field, materials having D-M interaction and/or single-ion anisotropy, the MD coupling further influenced by spin-orbit coupling and crystal field effects \cite{suzuki2008magnetodielectric, tackett2007magnetodielectric}. These factors collectively shape the MD properties, offering new avenues for tuning material behavior in response to electric and magnetic fields.


The transition metal phosphorus trichalcogenides MPX$_3$  (M = transition metal, X = S, Se) represent a family of compounds that exhibit two-dimensional characteristics in both their chemical structure and for certain metals (M = Mn, Fe, Co, Ni). These compounds form layered structures where the layers are weakly bonded by vdW interactions. Interestingly, the magnetic ground states of the MPX$_3$ family exhibit variations depending on the transition metal element. The ground state spin arrangement exhibits either N\'eel or zigzag depending on the exchange coupling energies between transition metal ions. These structures exhibit Heisenberg (MnPS$_3$), XY (NiPS$_3$), or Ising (FePS$_3$) behavior, determined by interactions between the first-, second-, and third-neighbor metal atoms. Although interlayer exchange coupling is relatively weak, it significantly influences the magnetic order along the \( c \)-axis. The variety of spin dimensionality and spin orders make this family of crystals attractive to study magnetic and structural interplay in bulk as well as in atomically thin layers. 

All members of the family MPS$_3$ share the monoclinic space group \( C2/m \)~\cite{ouvrard1985structural}. Within these structures, the transition metal ions arrange in a planar honeycomb lattice, each enclosed within an octahedron formed by sulfur atoms at its vertices, while pairs of phosphorus atoms occupy the centers of the hexagonal units. For example, in FePS$_3$, the magnetic Fe atoms are arranged in a honeycomb lattice, with each Fe atom octahedrally coordinated by six S atoms, exhibiting a slight trigonal distortion. P atoms are bonded to three S atoms and another phosphorus atom, forming \([P_2S_6]^{4-}\) unit. These sulfur layers are stacked along the \(c\)-axis in an ABCABC sequence, resulting in a monoclinic structure with the space group \(C2/m\) \cite{xie2019crystallographic}. The lattice parameters at 2 K temperature were measured \cite{lanccon2016magnetic} as \textit{a} = 5.94(4) \AA{}{}, \textit{b} = 10.26(2) \AA{}{}, \textit{c} = 6.60(6) \AA{}{}, and $\beta$ = 108.3(7)$^\circ$.

 An antiferromagnetic order develops in FePS$_3$ around the N\'eel temperature ($\sim$118 K) with strong anisotropy. It exhibits Ising-type antiferromagnetic ordering, where the high-spin (S = 2) Fe$^{2+}$ moments are predominantly aligned along the out-of-plane direction \cite{xie2019crystallographic}. In contrast, NiPS$_3$ undergoes an XY-type transition with T$_N$ = 155 K, while MnPS$_3$ shows a Heisenberg-type transition with T$_N$ = 78 K. The magnetic structure of FePS$_3$ has propagation vector \( \mathbf{k}_M = \left[0 \ 1 \ \frac{1}{2} \right] \) \cite{lanccon2016magnetic}. For FePS$_3$ and other members of the MPX$_3$ family, a detailed summary of their crystal characteristics and magnetic properties is provided in Table I.

Before going to the details of MD coupling results in representative 2D magnetic materials, we first present microscopic mechanisms for the appearance of MD coupling. One possible mechanism for having MD properties is spin-phonon coupling.


\begin{table*}[htb]
\centering
\caption{Comparison of crystal structure and magnetic properties of selected antiferromagnetic (AFM) 2D MPX$_3$ (M = Fe, Co, Ni, Mn, Cr; X = S, Se), 2D metal oxyhalide materials, RuCl$_3$ and 2D multiferroics.}

\resizebox{\textwidth}{!}{
\begin{tabular}{|c|p{2.5cm}|p{6cm}|p{3cm}|c|p{4cm}|c|}
\hline
\textbf{Material} & \textbf{Crystal Symmetry} & \textbf{Lattice Parameters} & \textbf{Magnetic Propagation Vector (k$_M$)} & \textbf{N\'eel Temperature (T$_N$)} & \textbf{Spin Dimensionality \& Spin Order} & \textbf{References} \\ \hline
FePS$_3$ & C2/m & a = 5.94(4) \AA{}, b = 10.26(2) \AA{}, c = 6.60(6) \AA{}, $\beta$ = 108.3(7)$^{\circ}$ & [0 1 $\frac{1}{2}$] & $\sim$118 K & Ising, zigzag & \cite{nauman2021complete,lanccon2016magnetic} \\ \hline
NiPS$_3$ & C2/m & a = 5.8120 \AA{}, b = 10.0700 \AA{}, c = 6.6320 \AA{}, $\beta$ = 106.9800$^{\circ}$ & [0 1 0] & $\sim$155 K & XY, zigzag & \cite{joy1992magnetism,chandrasekharan1994magnetism,lanccon2018magnetic,shemerliuk2021tuning,wildes2015magnetic,brec1986review} \\ \hline
CoPS$_3$ & C2/m & a = 5.9010 \AA{}, b = 10.2220 \AA{}, c = 6.6580 \AA{}, $\beta$ = 107.1700$^{\circ}$ & [0 1 0] & $\sim$122, $\sim$132 K & Heisenberg, zigzag & \cite{lanccon2016magnetic,wildes2017magnetic,brec1986review,mayorga2017layered} \\ \hline
MnPS$_3$ & C2/m & a = 6.0770 \AA{}, b = 10.5240 \AA{}, c = 6.7960 \AA{}, $\beta$ = 107.3500$^{\circ}$ & [0 0 0] & $\sim$78 K & Heisenberg, N\'eel & \cite{rule2002contrasting,masubuchi2008phase,joy1992magnetism,brec1986review,kurosawa1983neutron,ressouche2010magnetoelectric} \\ \hline
FePSe$_3$ & R-3 & a = 6.262 \AA{}, c = 19.805 \AA{} & [$\frac{1}{2}$ 0 $\frac{1}{2}$] & 119 $\pm$ 1 K & Ising, zigzag & \cite{wiedenmann1981neutron,bhutani2020strong} \\ \hline
NiPSe\textsubscript{3} & C2/m & a = 6.1604 \AA{}, b = 10.6768 \AA{}, c = 6.9323 \AA{}, $\beta$ = 107.1791$^{\circ}$ & --- & $\sim$206 K & Heisenberg, zigzag & \cite{le1982magnetic} \\ \hline
MnPSe$_3$ & R-3 & a = 6.387 \AA{}, c = 19.996 \AA{} & [0 0 0] & 74 $\pm$ 2 K & XY, N\'eel & \cite{wiedenmann1981neutron,bhutani2020strong} \\ \hline
CrPSe$_3$ & C2/m & a = 6.1120 \AA{}, b = 10.6950 \AA{}, c = 6.6950 \AA{}, $\beta$ = 107.770$^{\circ}$ & [0 0.452 0.215] & $\sim$126 K & Heisenberg, zigzag & \cite{baithi2023incommensurate} \\ \hline
CrOCl & Pmmn & a = 3.86 \AA{}, b = 3.20 \AA{}, c = 7.72 \AA{} & [0 $\frac{1}{2}$ 0] & $\sim$15 K & Stripy, N\'eel & \cite{angelkort2009observation,qing2020magnetism} \\ \hline
CuCrP$_2$S$_6$ & C2/c & a = 5.9098 \AA{}, b = 10.2447 \AA{}, c = 13.3644 \AA{}, $\beta$ = 106.97$^{\circ}$ & --- & $\sim$30 K & Heisenberg, N\'eel & \cite{park2022observation,hong2024structural,selter2023crystal} \\ \hline
\end{tabular}
}
\end{table*}


\subsection{Spin-Phonon Coupling}
\label{sec3.1}   
The dielectric properties of a material depends on optical phonon frequencies which is described by Lyddane-Sachs-Teller relation \cite{barker1975long}.  Magnetic fluctuations can influence optical phonon frequencies through spin-phonon coupling. While phonon frequencies are generally unaffected by magnetic properties, materials with strong spin-phonon coupling experience changes in phonon frequencies due to magnetic fluctuations, which subsequently impact the dielectric response. Consequently, magnetic field-induced modifications to these fluctuations contribute to the MD effect through this coupling mechanism. 

Therefore, the function g(q) represents the interaction between phonons and magnetic order, capturing the influence of spin fluctuations on lattice vibrations and the dielectric properties of the material.

The function \( g(q) \) physically represents the interaction between phonons and magnetic order. Lawes {\it et al.} \cite{lawes2003magnetodielectric} suggested that the functional form of \( g(q) \) can be derived by expanding the magnetic exchange integral with respect to atomic displacements. It is expressed as:

\begin{equation}
    g(q) = \gamma (1 - \cos(qR)),
\end{equation}

where \( \gamma \) is the coupling constant, and \( R \) denotes the displacement between nearest neighbors. This expression is a valid approximation for long-wavelength \( q \)-modes.  Notably, \( g(q) \) attains its maximum value near the Brillouin zone boundary.

\begin{figure}
\includegraphics[scale=1]{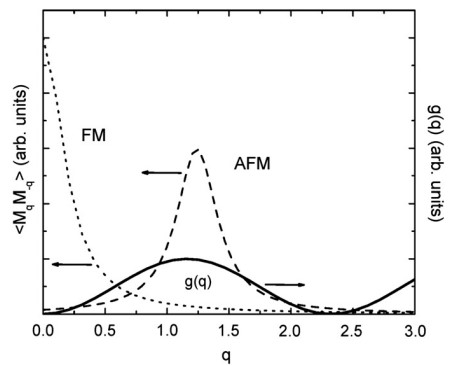}
\caption{Schematic curves showing MD coupling costant g(q) by the solid line. The dotted line and the dashed line showing spin-spin correlation functions for FM and AFM orders, respectively \cite{lawes2009magnetodielectric}. Reprinted from \cite{lawes2009magnetodielectric}, Copyright (2019), with permission from Elsevier.
\label{Figure1}}
\end{figure}

This correction of the basic model is adequate for examining MD coupling at AFM ordering transitions. Although \(M\) is zero in an AFM, the correlation function \(\langle M_q M_{-q} \rangle\) will exhibit a peak near the magnetic Bragg vector at the zone boundary as \(T\) \(\sim T_N\) \cite{lawes2003magnetodielectric}. And for FM materials that peak appears at \(q\) = 0. In the paramagnetic region,  there exists a fixed correlation between spins, which manifests as a slight shift in the dielectric constant. In the ferromagnetic phase, this correlation diminishes to zero, leading to a small discontinuity in \( \varepsilon \) around \( T_C \). For antiferromagnetic materials, the magnetodielectric effect is best explained by Fig. \ref{Figure1}, where \( g(q) \) reaches its maximum near the zone boundary. If \( g(q) \) and \( \langle M_q M_{-q} \rangle \) overlap (as observed in antiferromagnetic systems), a strong magnetodielectric effect is observed. Conversely, if these terms do not align, the magnetodielectric effect diminishes to zero.

 
\setlength{\parskip}{0pt}


\begin{figure}
\includegraphics[width=\linewidth]{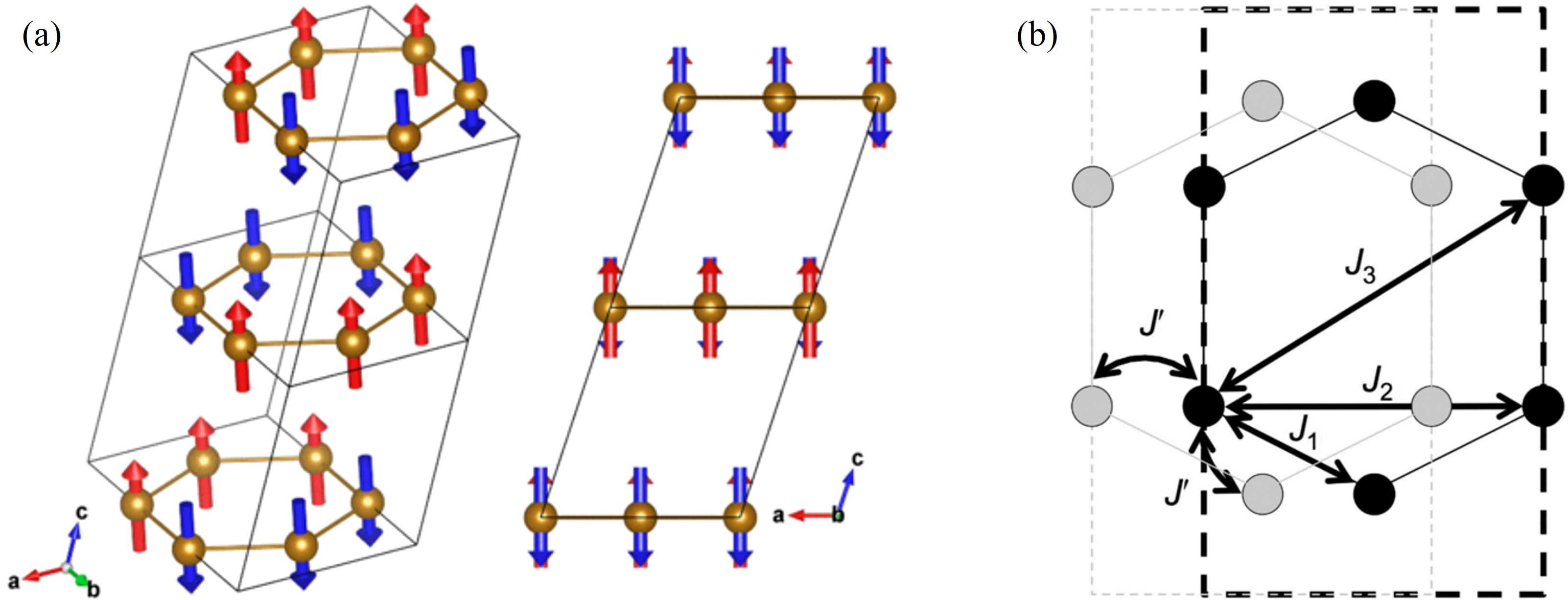}
\caption{ (a) The magnetic structure of FePS$_3$ shown in two different orientations. The magnetic moments of each Fe ion has magnitude of 4.52 $\pm 0.05\ \mu_{B}$ and are color-coded to show the difference between two different moment directions. (b) The interplanar and intraplanar exchange parameters between Fe ions. The black and grey cicles represent Fe ions in the ab basal plane and in the plane displaced along \textbf{c} by one lattice unit \cite{lanccon2016magnetic}. Reprinted (figures) with permission from \cite{lanccon2016magnetic}, Copyright (2017) by the American Physical Society.  
\label{Figure2}}
\end{figure}
\setlength{\parskip}{0pt}


 
In this review, we will be focused on the spin-phonon coupling phenomena for 2D magnetic materials. The mutual interactions between spin and lattice degrees of freedom have been found to exist in these materials down to few layer thickness. Mostly, the spin-phonon coupling in these materials has been identified through Raman measurements by analyzing phonon energies and lifetimes which diverge from the anharmonic behavior below magnetic transitions \cite{ghosh2021spin}. Direct visualization of the coupling between spin and lattice degrees of freedom through dielectric measurements with magnetic field has little explored in these layered materials. 

\begin{figure*}
\includegraphics[scale=0.75]{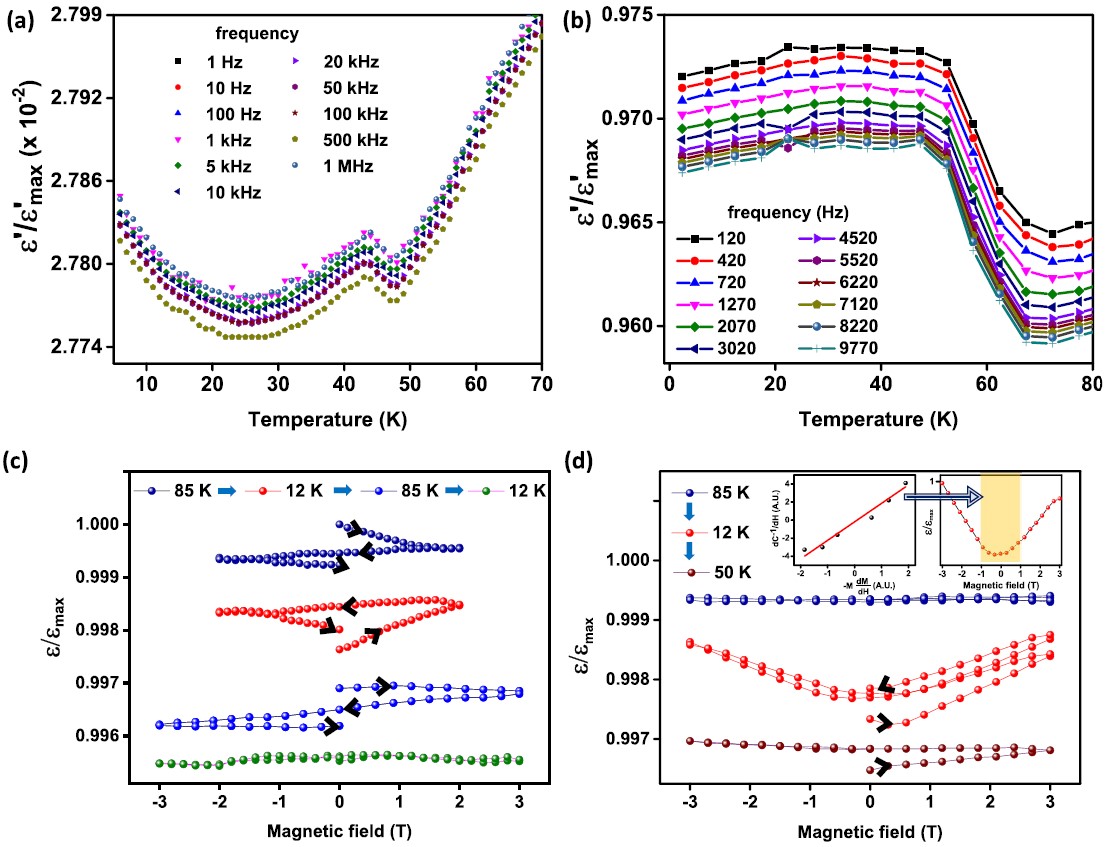}
\caption{ Temperature dependent (a) out-of-plane and (b) in-plane normalized dielectric constant at some selected frequencies. The change in the dielectric constant $sim$ 50 K show a prominent jump along in-plane direction. Magnetic field variation of (c) out-of-plane and (d) in-plane dielectric constant at some selected temperatures. The data were taken at 85 K and then at 12 K temperatures within $\pm$ 2T magnetic field and then repeats for $\pm$ 3T magnetic field for out-of-plane case. For in-plane measurements, the data were taken at 85 K, then at 12 K and finally at 50 K temperatures by varying magnetic field between $\pm$ 3T \cite{ghosh2023anisotropic}. Reprinted (figures) with permission from \cite{ghosh2023anisotropic}, Copyright (2017) by the American Physical Society.  
\label{Figure3}}
\end{figure*}

Among the various 2D magnetic materials, FePS\(_3\) stands out as a model AFM insulator as the magnetic ordering temperature remains the same down to monolayer limit \cite{lee2016ising}. Detailed MD coupling have been studied on its in-plane and out-of-plane directions \cite{ghosh2023anisotropic}.
Through temperature dependent in-plane and out-of-plane dielectric studies an indication of the influence of spin-phonon coupling has been realized in FePS$_3$. Apart from that, the low-temperature dielectric data reveals a frequency-independent anomaly in the dielectric constant $\epsilon$$^{\prime}$ around $sim$ 50 K, characterized by a sudden jump in both the out-of-plane [see Figure \ref{Figure3} (a)] and in-plane [see Figure \ref{Figure3} (b)] geometries.  Notably, the characteristic Raman modes exhibit an unusual downturn below $sim$ 50 K in the deviation from phonon anharmonicity (\(\Delta \omega\)) and the DC susceptibility ($\chi$$_{dc}$) displaying an unusual upturn (see Fig. 1. in \cite{ghosh2023anisotropic}).
The anomaly in the dielectric constant is typically associated with magnetic phase transitions \cite{lawes2003magnetodielectric, kimura2003magnetocapacitance, park2010effect} or ferroelectric ordering \cite{schrettle2008switching, schrettle2011relaxor, shi2020antiferromagnetic}. Another possibility of arising such anomalies can be associated with magnetic field-induced quantum fluctuations in an AFM at low temperatures \cite{ono2011magnetic}. However, the above mentioned possibilities have been denied through detailed analysis and discussion. For the possible explanation of dielectric anomaly predominantly along in-plane measurements, dielectric response in FePS\(_3\) with a magnetic field (\(H\)) applied parallel to the \(c\) axis, and in-plane direction have been checked.
\\

The effect of magnetic field on the dielectric response have been checked at a frequency of 100 kHz to avoid space-charge artifacts. In the out-of-plane case, the effect of the applied magnetic field on the dielectric response varies significantly above and below 50 K temperature (temperature where dielectric anomaly starts to appear). Figure \ref{Figure3}(c) shows the change in dielectric permittivity as the magnetic field is gradually swept between 0 T and \(\pm\) 2 T. At 85 K (navy), the permittivity decreases continuously with increasing and decreasing magnetic field, showing a \(\sim-0.08\%\) change after one complete cycle.

Lowering the temperature to 12 K (red) results in a marked change in dielectric response, with permittivity initially increasing rapidly from 0 T to \(2\) T, then decreasing from \(2\) T to \(-2\) T, and finally from \(-2\) T back to 0 T. The maximum change of  in capacitance is found \(\sim+0.08\%\). When the temperature is raised back to 85 K, the original curve shape reappears, even with the magnetic field increased to \(\pm\) 3 T. However, at 12 K, the change in permittivity becomes negligible (\(\sim 0.009\%\)), and the curve loops onto itself, losing the initial behavior observed in the virgin sample. This suggests that the magnetic field induces an irreversible change in dielectric permittivity at low temperatures. Below 50 K, temperature-induced structural frustration may cause micropolar regions to align differently than at 85 K, leading to moment locking that does not revert to the original state, even under a demagnetizing field.

For the in-plane case [see figure \ref{Figure2}(d)], the magnetodielectric response differs significantly from the out-of-plane scenario. No permanent locking effect is observed, and measurements taken consecutively at 85 K (navy), 12 K (red), 50 K (brown), and 12 K (not shown) show distinct MD coupling, most pronounced at 12 K (\(\sim+0.14\%\)) and diminishing with increasing temperature, becoming negligible at 85 K. This behavior may be attributed to the spin-phonon coupling in this system \cite{ghosh2021spin, vaclavkova2021magnon}.

Using crystal field calculations, the magnetic order in FePS$_3$ is shown to selectively couple to the trigonal distortions through partially filled t$_{2g}$ orbitals. Two unpaired spins in e$_{g}^{\pi}$ orbitals of FePS$_3$ allow efficient coupling between the magnetic order and trigonal distortions. This magnetoelastic effect is due to spin orbit coupling-mediated virtual transitions between e$_{g}^{\pi}$ and a$_{1g}$ orbitals \cite{lee2016ising,ergeccen2023coherent}.

According to Landau free-energy expansion, if P$^2$M$^2$ coupling is present in the material, which symmetry always allows, then $\frac{d(1/C)}{dH}$ should be proportional to M$\frac{dM}{dH}$ \cite{evans2015nature}. The first inset of figure \ref{Figure2}(d) shows the plot of $\frac{d(1/C)}{dH}$ versus -M$\frac{dM}{dH}$ at 12 K, forming a straight line between $\pm$ 1T magnetic field, but deviating beyond this range.

As the temperature decreases, lattice parameter distortion (with the a axis contracting and the b axis expanding \cite{evans2015nature, lanccon2016magnetic, budniak2022spectroscopy, jernberg1984feps3}) combined with anisotropy leads to complex domain interactions, causing frustration and freezing below $\sim$50 K. This explains the significant temperature shift in $\chi$$^{\prime}$ peaks and the unusual jump in the dielectric spectrum. Such behavior could result in domain-wall motion or related dynamics at low temperatures, governed by anisotropy constants \cite{nauman2021complete}. The anomalous $\chi$$^{\prime}$ response, with two sets of frequency-dependent peaks, may indicate multiple domain-wall phenomena. Similar temperature-induced domain-wall movement has been observed in other Ising systems like CoNb$_2$O$_6$ \cite{sarkis2021low}. 

To understand the magnetodielectric measurements in FePS\(_3\), we performed \textit{ab initio} calculations on the bulk system. The Fe ions, arranged in a honeycomb lattice, exhibit AFM-z ordering. XRD studies have shown deviations in the in-plane lattice constants from hexagonal symmetry \cite{murayama2016crystallographic} and nonequivalent Fe-S bond lengths, indicating in-plane crystallographic anisotropy \cite{geraffy2022crystal}. This suggests that the AFM-z phase within the monolayer has a preferred orientation \cite{amirabbasi2023orbital}.


\begin{figure}
\includegraphics[width=\linewidth]{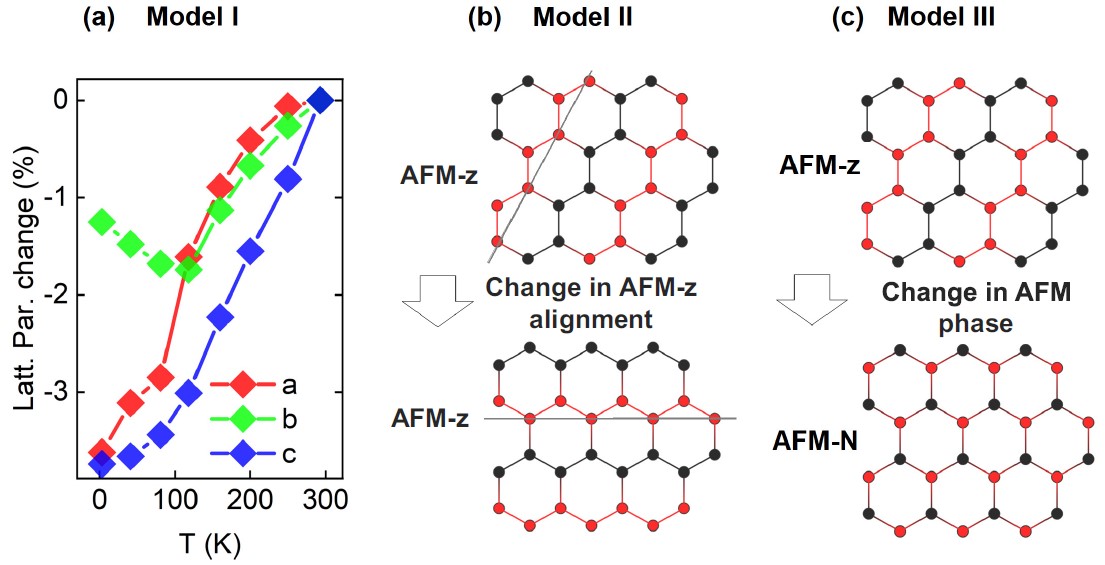}
\caption{ The different factors behind the dielectric anomaly are examined. (a) In Model I, the anomalous behavior of in-plane lattice parameters are examined.  Change in lattice parameters with temperature have been adopted from \cite{murayama2016crystallographic}. (b) In models II and III, two different possibilities of changing AFM-z orientation are assumed during calculation \cite{ghosh2023anisotropic}. Reprinted (figures) with permission from \cite{ghosh2023anisotropic}, Copyright (2017) by the American Physical Society. 
\label{AG_Fig4f}}
\end{figure}
\setlength{\parskip}{0pt}


To explain the significant dielectric jump around 50 K in the in-plane geometry, we considered three models: (I) changes in lattice parameters with temperature \cite{murayama2016crystallographic}, (II) orientation changes of the AFM-z phase within the monolayer, and (III) the impact of magnetic phases (AFM-z and AFM-N). We focused on the in-plane dielectric contributions, as they are five times larger than the out-of-plane ones.

In model I, lattice parameter changes led to minor dielectric variations, with a noticeable ionic contribution kink at 80 K for U = 5.3 eV. 
Two benchmark values equal to U=2.6 eV and U=5.3 eV was used for the calculations corresponding to the two extrema. Detailed discussion on the results is tabulated in table S1 (see the supplementary information of \cite{ghosh2023anisotropic}.

Model II, involving changes in AFM-z orientation, resulted in a larger relative dielectric contribution $\Delta$$\epsilon$$_{ion}$ (0.6\%) compared to model I (0.3\%-0.5\%). 

Model III revealed that the strongest in-plane dielectric changes (up to 8\%) occurred due to transitions between AFM-z and AFM-N phases, but these results did not match the observed experimental jump, excluding this model as the cause.

The preferred orientation of the magnetic alignment, influenced by structural in-plane anisotropy and lattice parameter changes, likely contributes to the dielectric jump at 50 K. The relative change in magnetic alignment, particularly in model II, aligns with the experimentally observed shift (\(\Delta \epsilon_0 = 0.8\%\)). This suggests that the dielectric anomaly around 50 K arises from the reorientation of the AFM-z phase within the plane, facilitated by in-plane structural anisotropy.

\subsection{Extrinsic origin of MD coupling}

The magnetodielectric (MD) effect arises from the intricate coupling between magnetic and dielectric properties in materials. This coupling can originate from both intrinsic and extrinsic mechanisms, which must be carefully distinguished to fully understand and exploit MD materials.

Intrinsic MD coupling arise from fundamental interactions between spins, charges and lattice vibrations in a magnetic material. Spin-lattice coupling is the major contributor, where the changes in magnetic order induce a change in optical phonon frequencies that we have discussed in the earlier section. 

Extrinsic contributions, on the other hand, arise from structural disorder such as defects, anion vacancies, dislocations, antisite disorder, grain boundaries, compositional off-stoichiometry and heterogeneities. In polycrystalline samples, extrinsic effects like Maxwell-Wagner polarization at grain boundaries can mimic intrinsic MD responses \cite{murthy2015investigation}. Similarly, thin-film materials exhibit strain-induced effects due to lattice mismatch with the substrate and interfacial phase with different crystal structure or electronic structures. which add extrinsic contributions to the overall MD properties. Therefore, impurity phases, anion vacancies, and surface contamination in nanoscale materials can further complicate the overall dielectric properties, particularly in the vicinity of magnetic transitions \cite{krishna2012magnetodielectric}.

To distinguish these effects, advanced experimental approaches are utilized, generally. To isolate intrinsic MD coupling from the extrinsic one high-quality single crystals or epitaxial thin films with minimal defects are used. Inelastic neutron scattering and Raman spectroscopy are the useful tools for studying spin-phonon coupling which is associated with intrinsic MD coupling. Spin-phonon interactions, where spin-spin correlations modulate phonon frequencies, provide a direct link between magnetic and dielectric coupling. Therefore, these studies are crucial to understand intrinsic mechanisms, as they reveal how lattice vibrations respond to the changes in magnetic order.

We investigated the MD coupling in high-quality FePS$_3$ crystals and flakes \cite{ghosh2023anisotropic}. Our study focused on the low-temperature region $\sim$50 K, where the interfacial and interlayer charge-related dielectric relaxations are negligibly small. Consequently, the extrinsic contributions to the observed MD coupling in FePS$_3$ are significantly reduced. In our earlier studies \cite{ghosh2021spin,vaclavkova2021magnon}, we reported the presence of strong spin-phonon coupling as well as magnon-phonon band hybridization, which are primordial to the intrinsic MD coupling in magnetic materials.


\section{Classification of Magnetodielectric Systems}
\label{sec3.2}
Several systems have been categorized under MD systems. These systems can be classified based on either spin-spin correlations or coupling between spin and any other primary order parameters (strain, polarization etc.). These systems include:

\begin{itemize}
    \item Frustrated spin systems
    \item Magnets with magnetoelastic coupling
    \item Magnets having spontaneous polarization 
\end{itemize}

In the above mentioned systems, dielectric properties have been shown to couple with magnetization and respond to applied magnetic fields. The resulting real and imaginary parts of dielectric properties influence and offer valuable insights into subtle changes in the spin configuration.     
In this review we started with the MD effect in simple ferromagnets (FM) and antiferromagnets (AFM), which are non-polar. Now we go to the above mentioned complicated frustrated and coupled spin systems. In this context, we frame recent advancements of MD coupling in 2D vdW materials.    

\subsection{Frustrated spin systems:}
\label{frustration}

In magnets with certain crystal structures such as the kagome, spinel and pyrochlore lattices, which are composed mainly of triangles or tetrahedra, the magnetic interaction between two spins often fail to produce a long-range order \cite{lawes2008magnetically, helton2007spin, mufti2011magnetoelectric, greedan2010geometrically}. In these systems, spin correlations develop below a temperature comparable to the magnetic interaction energy $\it{J}$ \cite{suzuki2005magnetocapacitance}. The disruption of long-range order arises due to their geometric constraints which results into disordered spins with spin fluctuation \cite{katsufuji2004magnetocapacitance}. Moreover, the spin fluctuation persists across a wide temperature range as an inherent characteristic. When such magnetic fluctuations interact with other degrees of freedom (such as charge), this coupling can be probed by electrical measurements other than direct spin measurements \cite{katsufuji2004magnetocapacitance, lawes2008magnetically, suzuki2005magnetocapacitance}. Thus, in magnetic insulators with similar coupling between fluctuating spins and charge degrees of freedom can be observed as a change in dielectric constant in response to applied magnetic field. 

In these materials, the change in dielectric properties upon application of magnetic field carry valuable information about the nature of magnetic exchange interactions \cite{saito2005magnetodielectric}. In some geometrically frustrated magnets, temperature-dependent MD have been found to deviate from the square of the magnetization, allowing the detection of magnetic fluctuations in their ground states \cite{katsufuji2004magnetocapacitance}.     

Other than triangular lattices in the above mentioned systems there are some other lattices which shows spin frustration. Recent studies have highlighted that chemical doping in some magnets [M$_2$Mo$_3$O$_8$, M: Mn, Fe, Co, Ni etc.] can significantly influence spin frustration and metamagnetic spin flop transitions \cite{morey2019n,yu2024crystal,kurumaji2015doping,abe2014magnetic,nakayama2011ferromagnetic}. Doping can alter spin orientations, leading to the stabilization of different magnetic states, such as ferrimagnetism or weak ferromagnetism. It can also enhance or modify magnetoelectric effects, such as increased linear and second-order ME responses or shifts in the spin easy axis, which triggers spin flop transitions when a magnetic field is applied \cite{kolodiazhnyi2011spin, yu2024crystal, tang2021metamagnetic}. These transitions are often accompanied by notable changes in electric polarization and dielectric constant, demonstrating the ability of doping to finely tune the magnetic and electric properties of these materials.

\begin{figure*}
\includegraphics[scale=0.5]{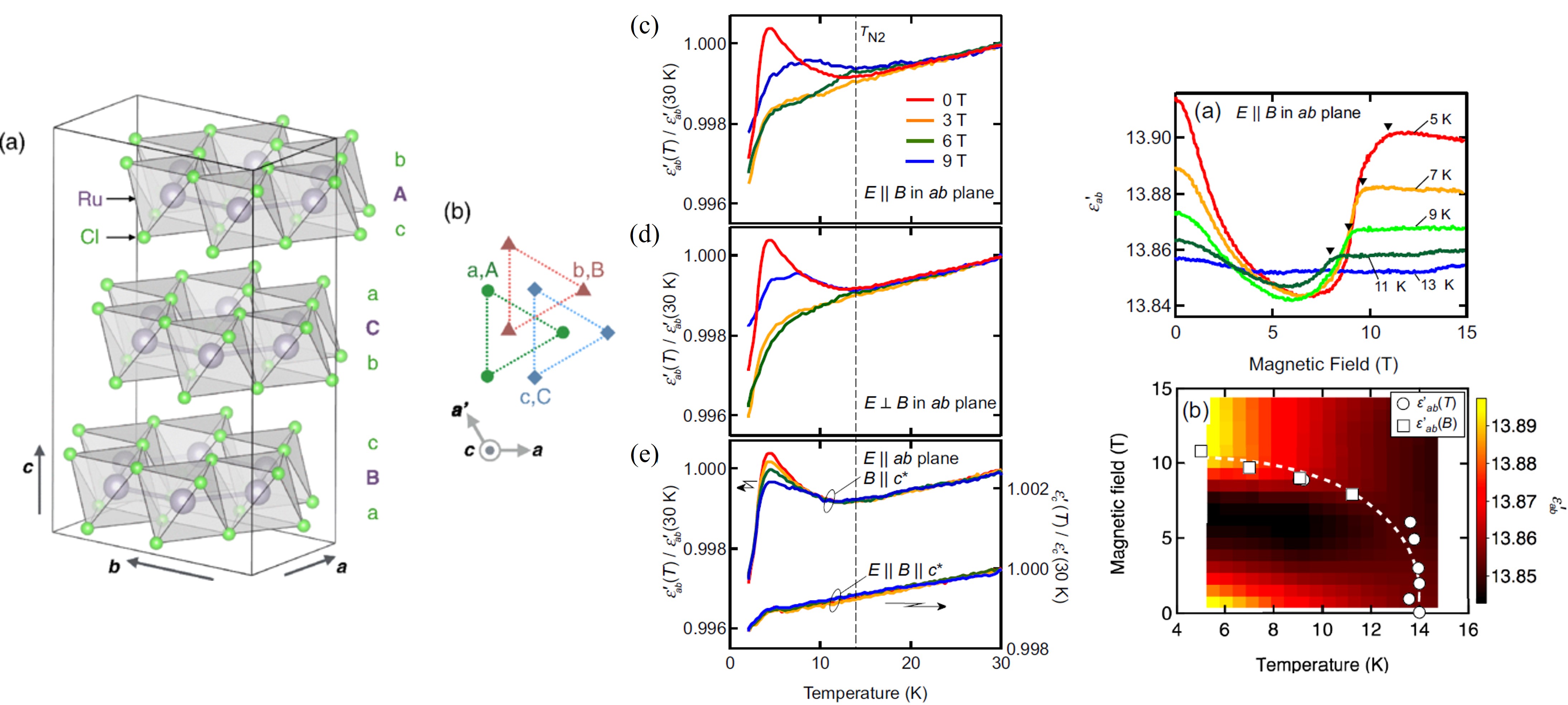}
\caption{Left (a) Structural representation of $\alpha$-RuCl$_3$ with monoclinic space group C2/m. Stacking indices for Ru honeycomb and Cl layers are expressed as capital (A, B, C) and small letters (a, b, c), respectively. left (b) Schematic view (from the top) of three successive layers of triangular sublattices on which Ru and Cl ions are located \cite{kim2016crystal}. Reprinted (figures) with permission from \cite{kim2016crystal}, Copyright (2016) by the American Physical Society.
The temperature variation of the dielectric constant measured at 100 kHz frequency at some selected magnetic fields in various directions. Electric fields and magnetic fields are applied along different directions with respect to the crystallographic axes. The dielectric spectra measured by applying (c) $E \parallel B$ in the ab plane, (d) E $\perp$ B in the ab plane, and (e) $E \parallel$ ab plane, $B \parallel c^{*}$, and $E, B \parallel c^{*}$. For better visualization of the changes with the applied magnetic fields, each data set has been scaled with the dielectric constant measured at 30 K. Right (a) Magnetic field variations of the dielectric constant measured by applying $E \parallel B$ in the ab plane at temperatures below \textit{T$_{N2}$}. Right (b) Phase diagram of the dielectric constant with temperature and magnetic fields measured by applying $E \parallel B$ in the ab plane. The circles and squares in the phase diagram are obtained from temperature and magnetic field dependent dielectric data \cite{aoyama2017anisotropic}. Reprinted (figures) with permission from \cite{aoyama2017anisotropic}, Copyright (2017) by the American Physical Society. 
\label{Rucl3}}
\end{figure*}
\setlength{\parskip}{0pt}



In many frustrated spin systems, the magnetically ordered state is stabilized through its interaction with lattice and/or charge degrees of freedom, which helps release the spin system's entropy \cite{robinson2011local}. \(\alpha\)-RuCl\(_3\) is such a 2D vdW  frustrated spin system. Though it is composed with honeycomb-lattice of magnetic Ru$^{3+}$ ions, it displays an unusual ground state due to significant amount of second and third neighbor exchange interactions or the presence of anisotropic exchange interaction \cite{kubota2015successive}. 

\(\alpha\)-RuCl\(_3\) has layered structure. Each individual \(\alpha\)-RuCl\(_3\) layers can slide over easily from one another due to weak vdW forces act between these layers. Depending on different interlayer stacking, \(\alpha\)-RuCl\(_3\) are found to possess different crystal symmetry such as rhombohedral $R\overline{3}$ \cite{glamazda2017relation}, trigonal \textit{P}3$_{1}$12 \cite{banerjee2016proximate, ziatdinov2016atomic} and monoclinic \textit{C}$2/\textit{m}$. High quality single crystals of \(\alpha\)-RuCl\(_3\) show a first order phase transition from monoclinic \textit{C}$2/\textit{m}$  symmetry \cite{johnson2015monoclinic, cao2016low} to the rhombohedral $R\overline{3}$ symmetry at  $\sim$ 150 K temperature \cite{namba2024two}. Crystals with stacking faults, however, do not show any structural transition. 

The magnetic properties of $\alpha$-RuCl$_3$ are significantly affected by its interlayer structure. The high quality single crystals show a single sharp magnetic transition at $T_N = 7$ K. By  introducing stacking faults, the N\'eel temperature (T$_N$) of $\alpha$-RuCl$_3$ raises from 7 K to 14 K by creating regions with a two-layer stacking sequence within the crystal \cite{cao2016low}.
This AFM ground state can be easily suppressed by an external magnetic field \cite{zheng2017gapless} and under pressure \cite{wang2018pressure}, resulting into novel quantum states. These findings highlight a strong coupling between lattice, spin, and orbital degrees of freedom.

The temperature profiles of \( \varepsilon^{\prime} \) are displayed in figure \ref{Rucl3}, revealing a distinct anisotropic MD response. In figure \ref{Rucl3}(e), when an electric field \( E \) is applied along the \( c^* \) axis, no significant temperature variation in \( \varepsilon^{\prime} \) is observed, and \( \varepsilon^{\prime} \) shows negligible magnetic field (\( B \)) dependence. Conversely, applying \( E \) along the \( ab \) plane at \( B = 0 \) T causes \( \varepsilon^{\prime}_{ab} \) to increase, following the Curie-Weiss law below \( T_{N2} = 14 \) K [figure \ref{Rucl3}(c) and figure \ref{Rucl3}(d)]. \( \varepsilon^{\prime}_{ab} \) broad peaks around 3 K before decreasing further with lower temperatures {\cite{aoyama2017anisotropic}.

The key observation is the nonmonotonic and anisotropic behavior of \( \varepsilon^{\prime}_{ab} \) under \( B \). When \( B \) is applied along the \( c^* \) axis [figure \ref{Rucl3}(e)], \( \varepsilon^{\prime}_{ab} \) remains largely unaffected, but applying \( B \) in the \( ab \) plane leads to a nonmonotonic change [figure \ref{Rucl3}(c) and figure \ref{Rucl3}(d)]. This significant response to in-plane \( B \) is due to the higher in-plane magnetic susceptibility compared to that along the \( c^* \) axis. As \( B \) is increased in the \( ab \) plane, \( \varepsilon^{prime}_{ab} \) initially decreases, with the peak at 3 K diminishing and disappearing above 3 T, only to increase again with further \( B \) enhancement. The anisotropy between longitudinal and transverse configurations in figure \ref{Rucl3}(c) and figure \ref{Rucl3}(d) is minimal. Though the peak in \( \varepsilon^{\prime}_{ab} \) and its dependence on \( B \) are reminiscent of magnetically induced ferroelectricity observed in some multiferroics, no significant macroscopic polarization was detected \cite{aoyama2017anisotropic}.

Figure \ref{Rucl3} right(a) illustrates the isothermal MD effect, highlighting the nonmonotonic variation in \( \varepsilon^{\prime}_{ab} \) as \( B \) is applied parallel to the \( ab \) plane. A downwardly convex MD effect is evident below 11 K, with a kink near 10 T at 5 K shifting to lower \( B \) as the temperature rises. This kink likely indicates the suppression of zigzag AFM order \cite{johnson2015monoclinic}. The temperature-magnetic field (\( T \)-\( B \)) phase diagram for \( \alpha \)-RuCl$_3$, derived from data in figure \ref{Rucl3}(c) and figure \ref{Rucl3} Right(a), is shown in figure \ref{Rucl3} Right(b) \cite{aoyama2017anisotropic}. 

So, the XY AFM state of $\alpha$-RuCl$_3$ has significant MD effect which was observed in the zigzag-type magnetic ordered state below 14 K, especially when electric and magnetic fields were applied within the honeycomb layer. The dielectric peak at lower temperature with strong frequency dependence, characteristic of relaxor antiferroelectrics where local polarization is induced by the zigzag-type magnetic order. These findings highlight the importance of considering both spin and lattice degrees of freedom to fully understand the ground state of $\alpha$-RuCl$_3$ \cite{aoyama2017anisotropic}.

AgCrS$_2$ and AgCrSe$_2$ are other 2D systems with magnetic frustration due to triangular planer lattice \cite{baenitz2021planar, damay2011magnetoelastic} whose dielectric properties are not explored much.


\subsection{Magnets having magnetoelastic coupling:}
\label{elastic}

  Magnetic frustration drives this structural phase transition \cite{yang2024magnetic, angelkort2009observation, zhang2023spin}. MD coupling on this system has found much larger at which spin flop transition takes place. However, detailed MD coupling studies are still required for CrOCl to understand the coupling at different magnetic phases.   

CrOCl is a layered material with an orthorhombic structure, where Cr ions occupy distorted octahedral sites coordinated by oxygen and chlorine atoms. It exhibits stripy antiferromagnetic (AFM) order below its N\'eel temperature (T$_N$ $\sim$14 K) \cite{gu2022magnetic}. In addition, a structural phase transition (from \(Pmmn\) space group the \(P2_1/m\) space group) occurs at the same temperature \cite{angelkort2009observation}. Magnetic frustration drives this structural phase transition \cite{yang2024magnetic, angelkort2009observation, zhang2023spin}. The AFM arrangement is characterized by spins aligning antiparallel within the layers, while interlayer interactions are weak, giving rise to its quasi-two-dimensional magnetic behavior \cite{angelkort2009observation}.

Magnetoelastic coupling in CrOCl is significant, arising from the interaction between magnetic order and lattice vibrations. Below T$_N$, lattice distortions occur to minimize the system's free energy, which leads to modifications in the exchange interactions mediated by the Cr-O-Cl super exchange pathways \cite{angelkort2009observation}. The spin anisotropy is primarily attributed to the crystal field environment of Cr$^{3+}$ ions and the spin-orbit coupling of chlorine \cite{rong2020ferromagnetism}. The low-dimensionality and strong spin-lattice coupling make CrOCl an interesting system for exploring the interplay of magnetism and lattice dynamics. Its response to external perturbations, such as strain or fields, highlights the potential to modulate its magnetic and electronic properties through mechanical or electronic coupling mechanisms \cite{schaller2023pressure}. These features place CrOCl within a broader class of materials exhibiting strong magnetoelastic and magnetoelectric phenomena, driven by the symmetry-breaking interactions inherent to their crystal and magnetic structures.

One intriguing feature of CrOCl is the appearance of a pyrocurrent, which emerges under specific experimental conditions, such as the application of magnetic fields or during temperature sweeps across the magnetic transition. The pyrocurrent is a result of changes in the material's polarization, which are linked to spin reorientation or magnetostriction effects associated with the onset of AFM order. This behavior underscores the coupling between magnetic and dielectric properties in CrOCl, suggesting a potential for magnetoelectric effects. MD coupling on this system has found much larger at which spin flop transition takes place. However, detailed MD coupling studies are still required for CrOCl to understand the coupling at different magnetic phases \cite{gu2023multi}.  

CrSBr is a 2D vdW material with a monoclinic crystal structure, exhibits A-type antiferromagnetic (AFM) ordering, characterized by ferromagnetic coupling within the layers and AFM interactions between them. Below the N\'eel temperature (T$_N$ $\sim$ 132-140 K), the magnetoelastic coupling arises due to the interplay of lattice vibrations and magnetic anisotropy \cite{bae2024transient,fei2024spin}. The latter is governed by anisotropic superexchange interactions mediated by bromine atoms. Bromine's spin-orbit coupling contributes to a pronounced magnetocrystalline anisotropy, which dominates over the single-ion anisotropy of Cr$^{3+}$ ions. In contrast, the large magnetoelastic coupling in FePS$_3$, mainly arises from the single-ion anisotropy of F(II). Structural distortions induced by magnetic ordering modulate the exchange pathways, reinforcing the magnetoelastic response. CrSBr exhibit strong spin-phonon coupling and multiple magnetic phases \cite{pawbake2023raman}, therefore, has enough scope for studying MD coupling.

Cr$_2$Ge$_2$Te$_6$ crystallizes in a rhombohedral structure with vdW bonding between layers. The material is a ferromagnet with a Curie temperature below T$_C$ $\sim$61 K. Below T$_C$ , magnetoelastic coupling is evident from spin-lattice interactions, where lattice distortions alter exchange interactions and magnetic anisotropy \cite{chen2022anisotropic,bazazzadeh2021magnetoelastic,watson2020direct}. The coupling is amplified by the layered structure, where the reduced dimensionality enhances spin-orbital-lattice interplay. Crystal field effects in this material are linked to the partially filled d-orbitals of Ge and Cr, which contribute to the observed coupling phenomena under strain or during lattice deformations \cite{zhao2022xgt}.

In all these materials, magnetoelastic coupling is closely tied to structural distortions and orbital effects. For instance, lattice vibrations modify the exchange integrals, while orbital-lattice interactions shift crystal field levels, reinforcing or altering the magnetic anisotropy. Such coupling effects are crucial for understanding the interplay between magnetic ordering and mechanical properties, especially in low-dimensional systems where reduced symmetry enhances these interactions.


The mechanical properties of these materials exhibit significant changes due to the onset of magnetic ordering, driven by the coupling between spin and lattice dynamics. Near the magnetic ordering temperature, materials such as FePS$_3$, CoPS$_3$, and CrOCl show measurable changes in mechanical parameters, including the resonance frequency and elastic constants. These anomalous behavior in mechanical properties are directly linked to the spin-lattice interactions that emerge during the onset of magnetic order. For example, in FePS$_3$ and CoPS$_3$, the lattice undergoes anisotropic distortions below the TN, which are evident in the compression and expansion of specific crystallographic axes \cite{houmes2023magnetic,pestka2024identifying}. These deformations indicate different elastic properties along different crystallographic axis below TN. Thereby spin correlations induce strong coupling with the lattice vibrations and the magnetic exchange interactions influences the material's mechanical response. In contrast, NiPS$_3$ exhibits relatively negligible changes, suggesting weaker magnetostrictive effects due to differences in its magnetic exchange pathways and orbital configurations \cite{houmes2023magnetic}.

Shear modulus studies provide additional insights into the interplay between magnetism and mechanical properties. In CrOCl and CrPS$_4$, for instance, the softening of the elastic constants near magnetic phase transitions highlights the strong spin-phonon coupling in these systems. The sensitivity of the shear modulus to magnetic ordering underscores its role as a critical parameter for detecting and characterizing magnetoelastic coupling \cite {li2022mechanical,houmes2024highly}.

Spin-frustrated materials, such as $\alpha$-RuCl$_3$, exhibit zigzag antiferromagnetic states, influenced by Kitaev-type interactions and Heisenberg exchange. These magnetic orders create anisotropic magnetostriction, where mechanical deformations align with spin configurations. In RuCl$_3$, the honeycomb lattice distortion is linked to strong magnetoelastic coupling, which becomes pronounced near the T$_N$. The interplay of lattice strain and magnetic fields results in anisotropic expansion and contraction, measurable via magnetostriction experiments \cite{kocsis2022magnetoelastic,schonemann2020thermal}. Measurements of thermal expansion and resonance frequency shifts provide insights into phase transitions, such as the gap structure associated with field-induced transitions

These observations collectively emphasize the intrinsic relationship between mechanical properties and magnetic order in 2D materials. As the magnetic order develops, the material's lattice dynamically adjusts to accommodate the spin configuration, leading to measurable changes in mechanical parameters. Such studies are crucial for advancing our understanding of magnetoelastic effects and their potential implications for the design of multifunctional devices that leverage the interplay between magnetic and mechanical phenomena.


\begin{figure*}
\includegraphics[scale=0.7]{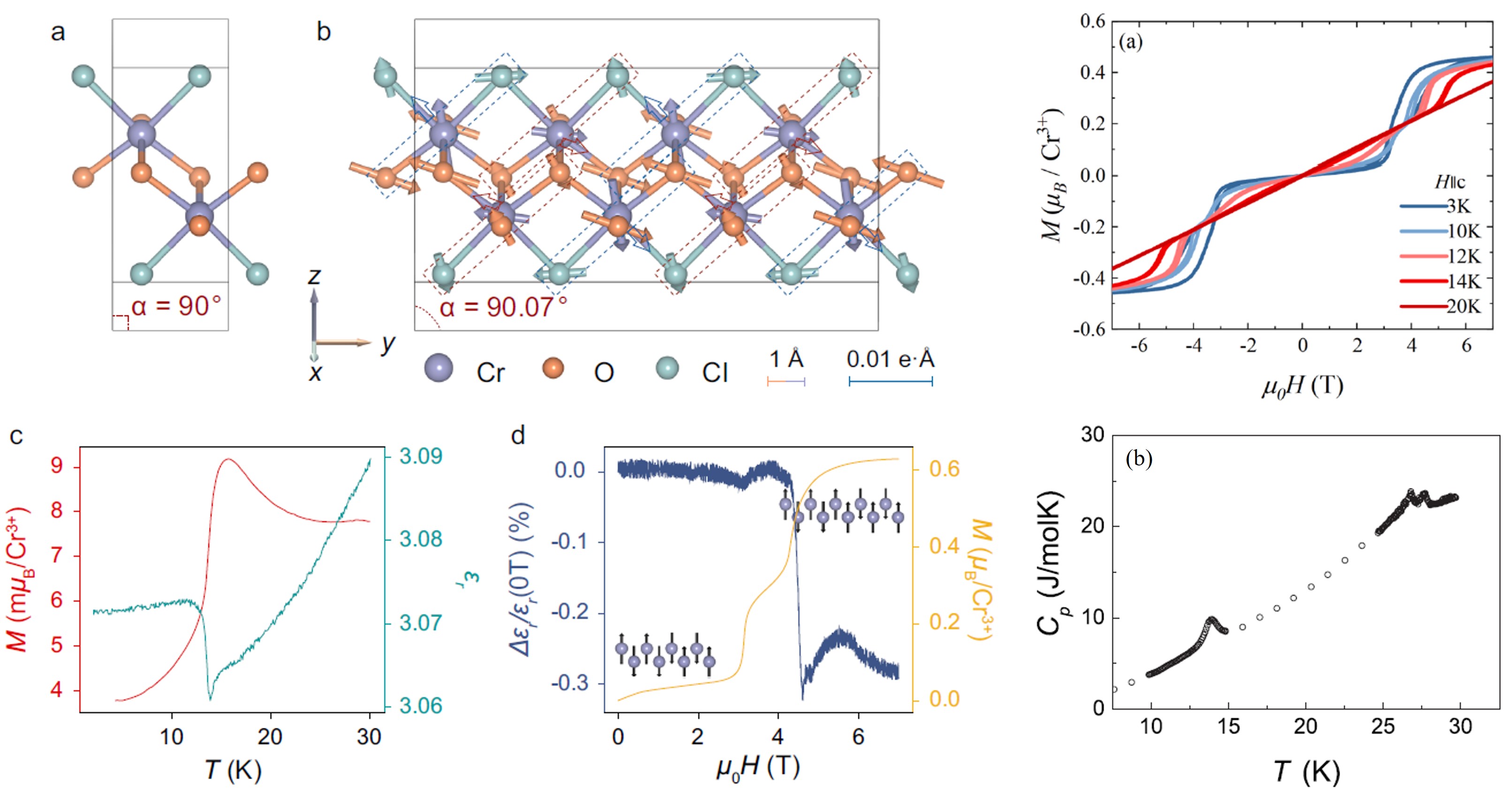}
\caption{\label{fig:epsart}Left (a) Crystal structure of CrOCl above the N\'eel temperature, characterized by the \(Pmmn\) space group within a unit cell. Left (b) Crystal structure of CrOCl below the N\'eel temperature, featuring the \(P2_1/m\) space group within a unit cell. The structural parameter \(\alpha\), which defines the crystal symmetry, is indicated in the unit cell. Vectors on each atom represent atomic displacements (magnified 100 times), while the dashed boxes and open arrows show the calculated antiferroelectric polarizations for different Cl-Cr-O chains. 
Left (c) Temperature dependence of the relative permittivity \(\varepsilon_r\) and magnetic moment of the bulk CrOCl crystal. 
\textbf{d} \(\Delta \varepsilon_r/\varepsilon_r(0\ T)\) plotted against the out-of-plane magnetic field, alongside the corresponding \(M-H\) curve at 2 K \cite{gu2023multi}. Reproduced from \cite{gu2023multi}. CC BY 4.0.
Right (a) The isothermal Magnetization with \( H \parallel c \) at selected temperatures \cite{yang2024magnetic}. Reprinted (figures) with permission from \cite{yang2024magnetic}, Copyright (2024) by Elsevier. 
Right (b) Temperature dependence of the specific heat \(C_p\) of CrOCl. Anomalies are distinctly observed at \(13.9 \pm 1\) K, and at 26.7 and 27.8 K \cite{angelkort2009observation}. Reprinted (figures) with permission from \cite{angelkort2009observation}, Copyright (2009) by the American Physical Society.} 
\label{crocl}
\end{figure*}
\setlength{\parskip}{0pt}

\subsection{Magnets having spontaneous polarization:}
\label{polar}

There are certain magnetic materials that exhibit spontaneous polarization, making them crystallographically unique. These materials are referred to as multiferroics, where both magnetic and ferroelectric orders coexist. Multiferroics are categorized into two classes based on the relationship between their magnetic  and FE ordering temperatures:

\begin{itemize}
    \item \textbf{Type I Multiferroics:} In these materials, the ferroelectric transition temperature (T$_{C}^{FE}$) is much higher than the magnetic ordering temperature (T$_{N}^{AFM}$). As a result, the coupling between magnetism and ferroelectricity is weak. The ferroelectricity in these systems often originates from structural distortions unrelated to the magnetic order. Examples include BiFeO$_3$, where ferroelectricity (T$_C$ $\sim$1100 K) dominates over antiferromagnetic ordering (T$_N$ $\sim$643 K), and PbVO$_3$ \cite{eerenstein2006multiferroic,ga2024electrically}.

    \item \textbf{Type II Multiferroics:} In these materials, the ferroelectric transition temperature (T$_{C}^{FE}$) is lower than the magnetic ordering temperature (T$_{N}^{AFM}$), indicating strong coupling between the two orders. Here, the ferroelectricity has a magnetic origin, often resulting from spin-lattice coupling or spin-induced electric polarization. Examples include TbMnO$_3$, where a helical spin order induces ferroelectricity below T$_C$ $\sim$28 K (with T$_N$ $\sim$41 K, NdCrTiO$_5$ and rare earth borates \cite {kajimoto2004magnetic,gautam2019symmetry,tripathy2023structural}
\end{itemize}

In Type I multiferroics, the primary focus is on leveraging their high-temperature stability for practical ferroelectric applications, such as in BiFeO$_3$-based devices. On the other hand, Type II multiferroics are of great interest for exploring strong magnetoelectric coupling, which can be utilized in magnetoelectric sensors and spintronic devices.
 
CuCrP$_2$S$_6$ (CCPS) is a vdW layered material that stands out as a promising multiferroic system, exhibiting multiferroicity down to a few atomic layers while maintaining air stability. Structurally, CCPS belongs to the monoclinic space group C2/c with lattice parameters approximately a = 5.920 \AA{}{}, b = 10.30 \AA{}{}, c = 13.38 \AA{}{}, and \textit{$\beta$}=106.7$^{\circ}$. The material undergoes a structural phase transition around 150 K, where Cu$^{+}$ ions exhibit antiferroelectric (AFE) ordering \cite{kleemann2011magnetic}. Additionally, below T$_N$ = 32 K, CCPS shows antiferromagnetic (AFM) ordering, involving collective interactions between magnetic Cr$^{3+}$ ions and off-center Cu$^{+}$ ions. 

Interestingly, below T$_N$, CCPS exhibits a magnetic-field-induced modulation of polarization perpendicular to the \textit{ab} plane, demonstrating magnetoelectric coupling \cite{park2022observation,susner2020temperature}. In the few-layer regime, emergent ferroelectricity and magnetoelectric effects have been observed, with nanoscale flakes displaying out-of-plane ferroelectric polarization even at room temperature, despite the absence of bulk ferroelectricity at this temperature. Experimental studies suggest that the coupling between magnetic moments and electric dipoles is evident from field-dependent magnetization measurements, revealing pinning effects likely due to magnetoelectric interactions. However, detailed structural insights into how the out-of-plane polarization couples with spin degrees of freedom remain elusive. These findings highlight the potential of CCPS in magnetoelectric and spintronic devices, leveraging its layered nature and coupling mechanisms

Nickel iodide (NiI$_2$) is a layered vdW material that displays type-II multiferroic behavior, where ferroelectricity is driven by a non-collinear spin structure rather than lattice distortions. Its trigonal crystal structure, belonging to the space group P$\bar{3}$m1, consists of triangular layers of Ni$^{2+}$ ions linked by iodine atoms. These layers stack along the c-axis, forming a 2D lattice.

As the temperature decreases, NiI$_2$ undergoes two distinct magnetic transitions. At approximately 76 K, it transitions into an interlayer antiferromagnetic (AFM) state, followed by a helimagnetic phase below 59.5 K. In the helimagnetic phase, spins form a spiral arrangement that breaks inversion symmetry, inducing ferroelectric polarization via the inverse Dzyaloshinskii-Moriya (DM) interaction. This coupling between spin and polarization demonstrates the rich multiferroic nature of NiI$_2$, where interlayer exchange interactions and magnetic anisotropy play significant roles in stabilizing these phases \cite{tokura2014multiferroics}.

In the monolayer form, NiI$_2$ exhibits a striped AFM order instead of the spiral phase observed in bulk, leading to antiferroelectric rather than ferroelectric properties. This emphasizes the importance of interlayer interactions in the bulk material, where they stabilize the spiral magnetic configuration necessary for multiferroic behavior \cite{blei2021synthesis}. Advanced techniques like second harmonic generation (SHG) provide valuable insights into symmetry-breaking processes, highlighting the interplay between magnetic ordering and polarization.

\begin{figure*}
\includegraphics[scale=0.75]{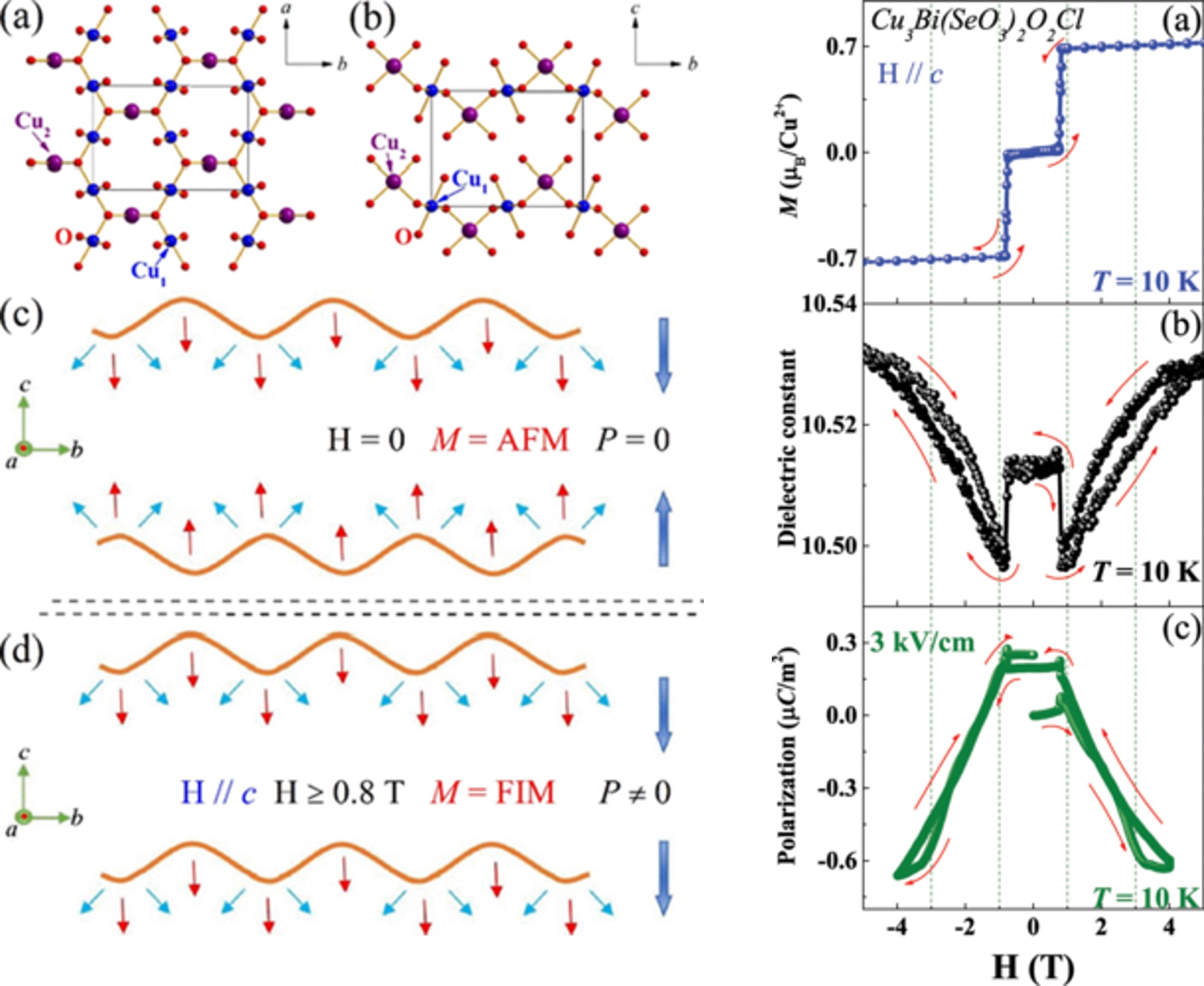}
\caption{Left (a) \textit{ab} and left (b) \textit{bc} plane of orthorhombic lattice (space group \textit{Pmmn}) of Cu$_3$Bi(SeO$_3$)$_2$O$_2$Cl. The small (blue), large (purple), and small (red) balls represent the Cu$_1$, Cu$_2$ and O atoms, respectively in the mentioned planes. Left (a) Schematic representation of layered spin structure in a \textit{bc} plane at \textit{H} = 0 (AFM state) and \textit{H} $\geq$ 0.8 T (ferrimagnetic like FIM state). Blue and red arrows indicate Cu$_1$ and Cu$_2$ spins, respectively. Right (a) Magnetic hysteresis loop, right (b) dielectric constant and right (c) polarization of the crystal with the variation of magnetic field along \textit{c} axis at a constant temperature 10 K. Red arrows indicate the measurement cycles \cite{wu2017anisotropic}. Reprinted (figures) with permission from \cite{wu2017anisotropic}, Copyright (2017) by the American Physical Society. 
\label{cbsoocl}}
\end{figure*}
\setlength{\parskip}{0pt}

 In a layered kagome system Cu$_3$Bi(SeO$_3$)$_2$O$_2$Cl (see figure \ref{cbsoocl}), a unique type-II multiferroic behavior has been found \cite{wu2017anisotropic}. The system crystallize into orthorhombic symmetry with \textit{Pmmn} space group. Figure \ref{cbsoocl}(a) and \ref{cbsoocl}(b) show the ab and bc planes of the orthorhombic lattice. The canted ferrimagnetic structure in the ab plane is illustrated in the schematic diagram of figure \ref{cbsoocl}(c). In this diagram, spin moments within each layer form the spin configuration, represented by solid orange lines. The spin structure is out of phase between layers. Within each layer, spin canting generates a finite Dzyaloshinskii-Moriya (DM) interaction vector, which induces electric polarization aligned with the spin moment, shown as large blue arrows in figure \ref{cbsoocl}(d).

 Cu$_3$Bi(SeO$_3$)$_2$O$_2$Cl (see figure \ref{cbsoocl}), magnetic field control of polarization appears below $\it{T_N}$ and in magnetic fields nearby spin flip transition at $\sim$ 0.8 T (see figure \ref{cbsoocl} Right(a)), where ferromagnetism and ferroelectricity coexist due to the magnetic field \cite{wu2017anisotropic}. Measurements show that this material has a strong anisotropic response in both magnetization and MD effects. The ferroelectricity triggered by spin-flip was confirmed by changes in electric polarization under a magnetic field, as shown in figure \ref{cbsoocl} Right(c). At the same magnetic field, the dielectric constant shows an abrupt jump and higher fields dielectric constant increases with magnetic field (see figure \ref{cbsoocl} Right(b)). This field-induced multiferroic effect is likely due to the breaking of twofold symmetry in the magnetic blocks, providing new insights into multiferroics in spin-flip metamagnetic materials.

\subsubsection{Interfacial and interlayer multiferroicity in vdW heterostructures:}
\label{Int_mult}

Materials with a lack of inversion symmetry do not exhibit spontaneous electric polarization. Similarly, materials that retain time-reversal symmetry do not exhibit magnetization. These fundamental symmetry constraints significantly limit the number of materials capable of simultaneously supporting electric polarization and magnetization, hence single-phase multiferroics \cite{taniyama2015electric}. 

Another effective approach to achieving multiferroicity is the design of artificial two-phase multiferroic heterostructures by ferromagnetic (FM) and ferroelectric (FE) layers. In such FM/FE heterostructures, at the interface, magnetic field controllable polarization or vice versa can be achieved \cite{maity2024engineering,panchal2018strain}. These materials have also exhibit MD properties due to spin-charge coupling. Additionally, the strain mismatch at the interface between substrate and layered thin films may induce dipole moment by structural distortions due to compressive or tensile strain. Similarly, at the interface of 2D vdW heterostructures ferroelectricity, magnetoelectric coupling and multiferroicity can be induced by strain mismatch, charge transfer, interlayer exchange interaction, interlayer sliding and band bending etc. \cite{zhang2024tunable,guo20222d,xu2024enhanced,liu2023tunable,feng2024van,arora2024ferroelectricity,maity2024electron}. These can be controlled by choosing suitable material for making vdW heterostructures and the interfacial phenomenon can be tuned by external electric and/ or magnetic fields. 
\section{Summary, conclusions and outlook}
\label{summary}

In this topical review, we have provided an overview of the coupled magnetodielectric (MD) properties in two-dimensional (2D) magnetic materials. We have explored various types of MD couplings, the necessary conditions for MD coupling, and briefly described different systems that exhibit this phenomenon.

The preceding sections of this review have shown that complex physical mechanisms such as spin frustration, magnetostriction, spin-lattice coupling, metamagnetic transitions, and structural anisotropy along with antiferromagnetic (AFM) ordering, significantly contribute to MD coupling when these mechanisms are closely linked to the charge degrees of freedom of that system. Moreover, their contribution to MD coupling does not necessarily depend on whether the material exhibits special symmetry, spontaneous polarization or linear magnetoelectric (ME) coupling. Rather, spin-phonon coupling is the most fundamental mechanisms for understanding MD coupling.

Materials that exhibit both the polarization and magnetization orders generally display much stronger MD coupling compared to others. In multiferroics, the polar mode that couples with magnetic ordering is typically soft, leading to a larger MD coupling. Generally, in multiferroics, when magnetic ordering breaks inversion symmetry, MD coupling becomes significant.

\subsection{Outlook}
\label{Outlook}

Despite over twenty years of comprehensive theoretical and experimental research on 2D vdW materials, they are not yet on the verge of being utilized in device applications. Moreover, our understanding of the coupling phenomena within the 2D framework is still in its early stages. Future research will undoubtedly focus on advancing suitable experimental probes for measuring nanoscale magnetism, magnetodielectric coupling and the hetero interface. Self-assembled structures, such as vdW heterostructures, represent a highly promising approach for developing new types of magnetoelectric materials.

In the future, vdW magnets, specifically A$_5$(TeO$_3$)$_4$X$_2$ (where A = Ni, Co, and X = Cl, Br, I), with their monoclinic structure in the C2/c space group, are expected to gain increasing attention due to their intriguing magnetic structures and variations in lattice symmetry. The crystal structure of this family, composed of three distinct A octahedra [A(1)O$_6$], A(2)O$_5$X], and A(3)O$_6$ arranged in a clamp-like corner-shared A$_5$O$_{17}$X$_2$] unit, offers promising avenues for future exploration in the study of magnetodielectric properties \cite{yu2024magnetic, pregelj2007magnetic, becker2007crystal}. 

Recently, Co$_5$(TeO$_3$)$_4$Cl$_2$ among vdW magnets A$_5$(TeO$_3$)$_4$X$_2$, where A = Ni, Co and X = Cl, Br, I) have found to exhibit ferroelectricity with magnetic field below T$_N$. These systems are expected to exhibit sufficiently high MD coupling. The concept of magnetic field induced ferroelectricity remains quite elusive on 2D materials at this stage. In-situ structural studies with magnetic field may ultimately result in a refined model of the ferroelectricity, serving as a foundational element for understanding MD coupling. 

Recently, 2D vdW CuCrS$_2$ and CuCrSe$_2$ have been found to grow in atomically thin layer and exhibit stable ferroelectricity up to very high temperatures \cite{zhou2024van, sun2024evidence, wang2024chemical}. These materials are classified as Type I multiferroics, as magnetic ordering develops at significantly lower temperatures \cite{tewari2024signature, tewari2014thermoelectric}. They are well-suited for studying MD coupling near magnetic transition temperatures.

2D rare-earth oxyhalides (REOXs: RE= La-Lu) present intriguing opportunities for both fundamental research and practical applications due to their novel properties \cite{holsa2000simulation}. Despite their promising potential, these are still in their early stages of exploration.  

Additionally, substantial efforts are focused on fabricating high-performance 2D vdW transistors using high-$\kappa$ dielectrics, with 2D LaOCl emerging as a particularly promising layered high-$\kappa$ dielectric according to density functional theory \cite{osanloo2021identification}. This suggests significant potential for its application in transistors \cite{si2018ferroelectric,dey2024negative}. Looking ahead, the exploration of magnetic properties in these 2D RE materials could unlock new opportunities in developing advanced magnetic devices and technologies.

These RE elements, with their distinctive 4f electron configurations, exhibit unique magnetic properties compared to transition metals. For instance, DyOCl single crystals show an A-type AFM structure, making them a promising candidate for realizing f-electron 2D magnetism (15).  The investigation of layered 4f magnets  also offers a valuable avenue for studying MD coupling.

The coupling between ferromagnetic and ferroelectric materials in oxide thin films have been largely investigated for MD coupling. However, MD coupling investigations in 2D vdW heterostructures are rare. MD effects in these heterostructure could reveal unexpected results as further experimental research is conducted on those. Recently, 2D multiferroic vdW heterostructure MnSe$_{2}$/In$_{2}$Se$_{3}$ has been found as potential candidates for nonvolatile memory device \cite{zhang2024electronic}. This type of vdW heterostructure, integrating one FE layer with another exhibiting FM or antiferromagnetic AFM order, holds significant promise for advancing MD coupling studies. Specifically, using CuInP$_2$S$_6$ as the ferroelectric insulator in such heterostructures could further enhance the exploration of MD coupling phenomena.


\section*{Acknowledgments}
The authors are thankful to the CSS facility at IACS and the facilities at UGC-DAE CSR Indore. Although the authors of this review led and executed the research, it was significantly enriched by the contributions of numerous individuals, including former members of the group and researchers with whom we have had close and long-standing collaborations over the years. We extend our deepest gratitude to Pradeepta Kumar Ghose, Suresh Bhardwaj, Sujan Maity, Somsubhra Ghosh, Satyabrata Bera,  Shibabrata Nandi. The authors are thankful to Vasant Sathe and his lab-members for their assistance with the Raman measurements. Authors gratefully acknowledge theoretical support from Magdalena Birowska, Milosz Rybak and the computing facilities of PL-Grid Polish Infrastructure for Supporting Computational Science in the European Research Space and of the Interdisciplinary Center of Modeling (ICM), University of Warsaw. K.D. acknowledges Dinesh Kumar Shukla for his constant support and guidance in instrumentation and measurements across various dielectric materials. K.D. and S.D. acknowledge support from the Technical Research Centre, IACS, Kolkata. A.G. acknowledges IISER-K for fellowship Financial support from IACS, DST-INSPIRE, and CSIR-UGC are greatly acknowledged. S.D. acknowledges financial support from DST-SERB under Grant CRG/2021/
004334 and Special Grant SCP/2022/000411.  

\section*{Data availability statement}
No new data were used in this study. 



\end{document}